B. Castillo López de Larrinzar[1], C. Xiang[2], E. Cardozo de Oliveira[2], N. D. Lanzillotti-Kimura[2,*], A. García-Martín[1,+]

# Towards chiral acoustoplasmonics

**Abstract:** The possibility of creating and manipulating nanostructured materials encouraged the exploration of new strategies to control electromagnetic properties. Among the most intriguing nanostructures are those that respond differently to helical polarization, i.e., exhibit chirality. Here, we present a simple structure based on crossed elongated bars where light-handedness defines the dominating cross-section absorption or scattering, with a 200% difference from its counterpart (scattering or absorption). The proposed chiral system opens the way to enhanced coherent phonon excitation and detection. We theoretically propose a simple coherent phonon generation (time-resolved Brillouin scattering) experiment using circularly polarized light. In the reported structures, the generation of acoustic phonons is optimized by maximizing the absorption, while the detection is enhanced at the same wavelength -and different helicity- by engineering the scattering properties. The presented results constitute one of the first steps towards harvesting chirality effects in the design and optimization of efficient and versatile acoustoplasmonic transducers.

## 1. Introduction

Objects are considered chiral when their mirror images cannot be superimposed, even after spatial rotations in three dimensions. Their relevance is such that they are critical in life systems since biologically relevant molecules are predominantly chiral (proteins, amino acids, etc.). [1] Handedness (right vs. left, dextro vs. levo) identifies the two different enantiomers in a chiral structure, and it is the crucial element determining how systems interact with their environment. [2]

Electromagnetic fields can be made to exhibit chirality, handedness, or helicity, as in the simplest case of circularly polarized light. Circularly polarized light can be used to probe and determine the chiral nature of molecules or other human-made structures. Normally, the chiral nature of the structure is reflected as quantitative differences in the values of the absorption or scattering cross-sections. Only recently, it has been put forward that it is possible to find helicity-dependent absorption resonances while the scattering remains essentially helicity-independent. [3] Our work demonstrates that, through the interactions between different elements, it is possible to make the absorption and scattering cross-sections radically and qualitatively different for the two circular polarizations. Even more, we show that the dominating cross-section can be switched from absorption to scattering by simply changing the polarization of the impinging beam.

The possibility of using plasmonic structures has recently appeared as a suitable tool in nanophononics. [4–12] For instance, the use of plasmonic metasurfaces to generate tunable acoustic phonons, the use of Swiss cross antennae to spatially probe strain distributions at the nanoscale, or the control of transmission of acoustic waves using tailored antennae are just examples of applications where plasmonics meets nanomechanics.

The main limitation in developing photonic applications based on plasmonic resonances is optical losses (a.k.a. absorption). Interestingly, the generation of coherent acoustic phonons is strongly dependent on the coupling of the incident light and, generally, how much it is absorbed. [4,5,7,8,13] Likewise, the capacity to detect coherent acoustic vibrations in plasmonic structures relies on the scattering cross-section. To maximize the performance it is necessary to comply with two conditions: maximum absorption and minimum scattering for excitation, and the reverse for detection. [14] These conditions are not usually found to occur simultaneously at the same wavelength.

Therefore, having handedness-dependent absorption and scattering cross-sections unlocks an invaluable tool to decouple the generation and detection of coherent acoustic phonons in nanostructures. The use of chiral acousto-plasmonic structures appears as a potential game-changer in developing a novel generation of phononic transducers. This is precisely the core of the present work, where we examine a simple chiral structure composed of two resonant interacting elements stacked vertically, [15] so that the chiral characteristics can be tuned via their mutual interactions. The system is yet maintained simple enough to understand the actual nature of the response observed, be open to future developments, and warrant fabrication for future experimental verification. We will show that this chiral plasmonic structure exhibits critical characteristics to be used as the basis of future ultrafast acousto-plasmonic devices, where light-handedness can be employed to discriminate generation and detection processes of co-herent acoustic phonons. In short, we seek to develop structures with the dominance of absorption or scattering in the cross-sections depending on the handedness of the incident radiation.

[1] Instituto de Micro y Nanotecnología IMN-CNM, CSIC, CEI UAM + CSIC, Isaac Newton 8, Tres Cantos, Madrid 28760, Spain
[2] Université Paris-Saclay, CNRS, Centre de Nanosciences et de Nanotechnologies, 10 Boulevard Thomas Gobert, 91120 Palaiseau, France
*email : daniel.kimura@c2n.upsaclay.fr
+email : a.garcia.martin@csic.es

The paper is organized as follows: in Section 2 we introduce the rotated bars platform as a fundamental yet simple chiral plasmonic structure and the chiral characteristics from the handedness-dependent optical response; in Section 3 we propose the use of these structures to generate and detect acoustic phonons, where circular polarization plays a key role in enhancing the signals; and finally, in Section 4, we present the conclusions and perspectives.

## 2. The chiral optical platform

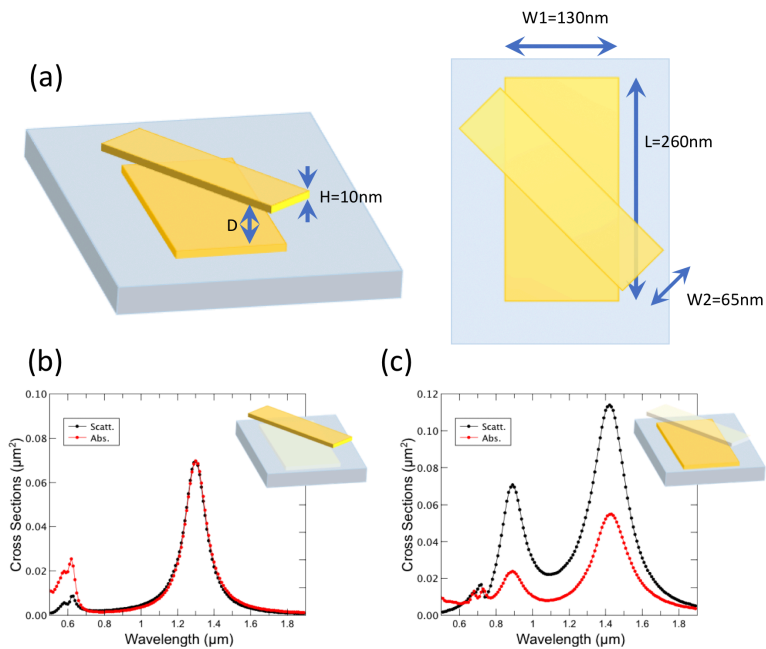

**Fig. 1:** (a) Sketch of the geometrical layout of the chiral structure depicting the dimensions of the bars (H, L, W1, W2) and the separation distance (D), the angle is set to 45 degrees. Scattering and absorption cross-sections for circularly polarized waves of the (isolated) top (b) and bottom (c) bars, showing that each bar leads to two different resonances per helicity.

The key idea is to have a system composed of as few elements as possible, with the constraint imposed by the fact that the mirror image cannot be superimposed with itself (i.e., geometrically chiral). Likely, the simplest version [16] of such a system is composed of two dissimilar metallic nanoplates with the long axis parallel to the polarization plane forming a given angle between them (45 degrees throughout this work) and vertically aligned perpendicular to the propagation direction of the impinging wave. As it is most usual in plasmonics, the metal to be employed is gold, and the dimensions of the plates are chosen so that each one of them separately has resonances in the Vis-NIR range of the electromagnetic spectrum, (see Fig. 1). The bottom bar, always in contact with a SiO$_2$ (n=1.456) substrate, has a thickness H=10nm, a length L=260nm, and a width of W1=130nm. The top bar has the same H and L, but the width is W2=65nm. This asymmetry in the width is introduced to mimic the profiles obtained from actual lithographic procedures. [17–19] In addition, it would allow fine-tuning over the spectral positions of the two resonances.

The numerical simulations have been performed using a commercial Finite Difference Time Domain (FDTD) solver for Maxwell's equations (Lumerical). The material Au properties are taken from the Lumerical's embedded database. [20] thus explicitly taking into account intrinsic losses and complex electromagnetic fields. We use an impinging plane wave with the proper polarization along the z-axis (perpendicular to the bars interaction axis) of unit amplitude in the whole simulation cell. The geometry of the cell (2.1 μm × 2.1 μm × 5 μm) ensures that perfectly absorbing boundary conditions do have a negligible effect on the electromagnetic fields obtained. We use a refined mesh in the nanostructures and in near field region (0.3 μm × 0.3 μm × 0.17 μm) of 1 nm × 1 nm × 0.75 nm, dx-dy-dz, respectively, growing uniformly up to a maximum of 15 nm out of the nearfield close to the simulation boundaries, so that convergence (to the best of our numerical capabilities) is attained. The refined mesh in the near field region equals a minimum of 500 points *per wavelength* in each direction. The total and scattered fields are then collected to give rise to the intensity color maps and the cross-sections. All fields are normalized to the amplitude of the impinging plane wave.

Figs. 1(a) and 1(b) present the scattering cross sections of each of the nanoplates in the absence of the other one, but in presence of the substrate, as a function of the wavelength under the incidence of a circularly polarized wave. As in the isolated case, none of the plates has any helicity dependence the two circular polarizations report the same results. The observed peaks represent the wavelengths for which a resonance of the individual nanoplate is excited, which are essentially four, corresponding to the long- and short-axis resonances in the pure helicity basis of circular polarization. Therefore, when the separating distance makes the two elements interact, the resulting scenario would be the complex mixing of four resonant modes, as pointed out in ref. [21] where a similar, somehow less anisotropic system, is decomposed in Mie multipoles.

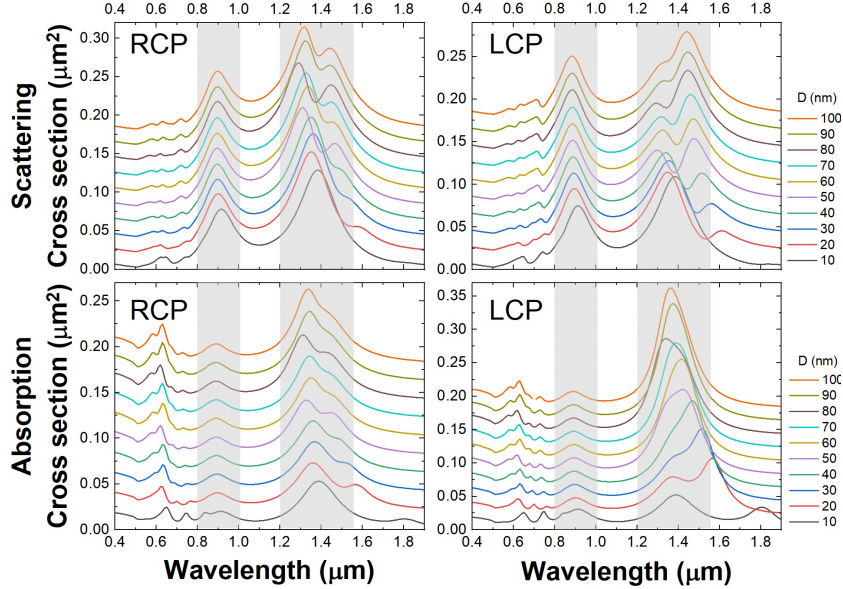

**Fig. 2:** Effect of the separation D of the bars (interaction) for the absorption (bottom) and scattering cross-sections (top) for the two helicities, left- and right-circular polarization, of the impinging plane wave.

When the two bars are superimposed, i.e. D=0 forming a planar system, the system is essentially a single anisotropic unit. Both absorption and scattering cross-sections (CS) for incoming Left Circular Polarized (LCP) or Right Circular Polarized (RCP) waves are expected to present two peaks, and the differences between the two polarizations are negligible. When the two bars are vertically displaced, D>0, the interaction between the near-field of each bar at resonance gives rise to hybrid states whose spectral position depends on the interaction (i.e. on D), thus allowing control over the overall response of the system in terms of CS. This effect is nicely depicted in Fig. 2, where we can observe two different resonant regions. The resonance located at ca. 875nm presents a clear dominance of absorption vs scattering for the two helicities, and no significant differences for LCP or RCP can be found. For the resonances located ca. 1.4µm there is a rich phenomenology due to interactions (see below). When the separation increases the field overlap decreases rapidly, [22,23] reaching a value from which the interaction can be regarded as negligible, and therefore the observed peaks closely represent the linear addition of the resonances of the two individual bars (see Fig. 1).

In Figure 3 we present selected curves extracted from Fig. 2 for vertical separations of (a) 10nm, and (b) 60nm. In (a) the separation is too short to appreciate significant differences between the two impinging helicities and we can see that scattering dominates clearly over absorption. However, in (b) the situation is radically different: for RCP there is a clear dominance of the scattering over absorption CS (close to a factor of 2), while for LCP the situation is reversed, absorption CS is predominant. Especially interesting is the point at ca. 1.36µm, where a sweet point can be observed in which the scattering CS for RCP is dominant and equivalent to the absorption CS for LCP, which is dominant for that polarization.

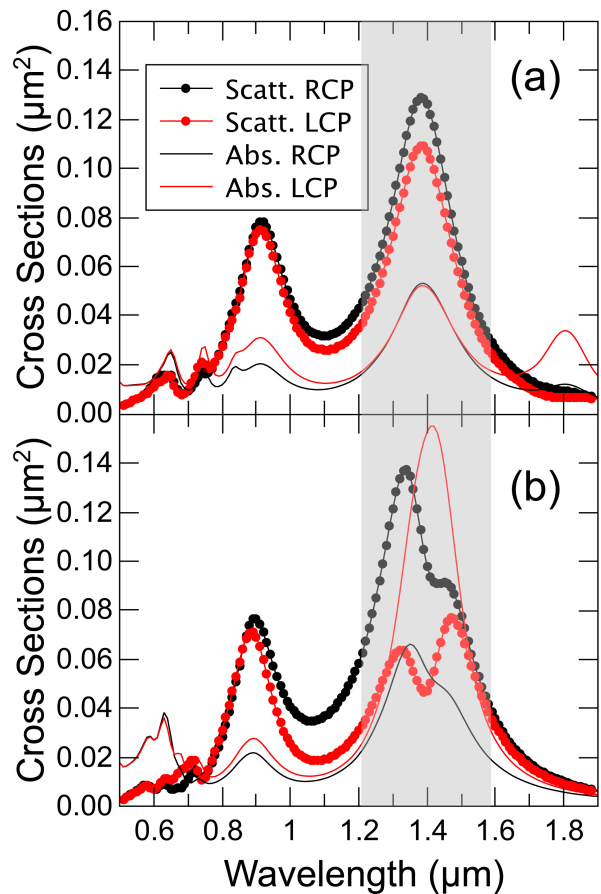

**Fig. 3:** Absorption and scattering cross-sections (structures over a substrate, but separated by air as in Fig. 2, for an edge-to-edge distance (a) 10nm and (b) 60nm. The grey shadow area is to reflect the frequency region where the effects take place.

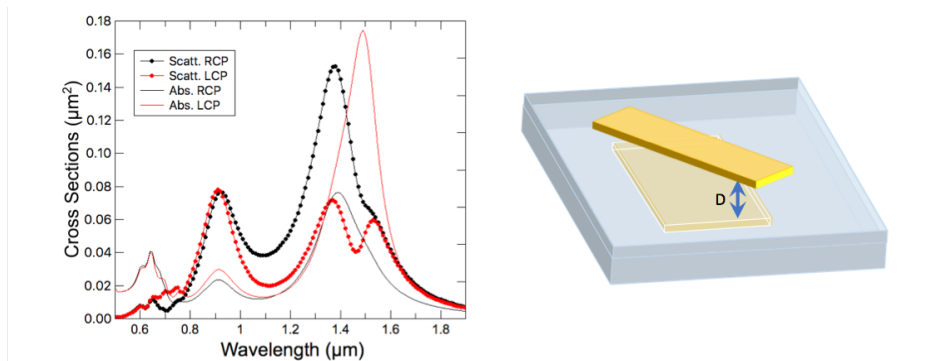

**Fig. 4:** Absorption and scattering cross-sections when the bottom bar is fully embedded in SiO2 and the top bar sits on the surface of SiO2, and the separation distance is D=80nm. As can be seen the same qualitative phenomenology as in Fig. 3 is obtained.

So far we have presented the phenomenon we were looking for, but with air between the bars. This configuration allowed us to keep the geometry and environment as simple as possible. However, the experimental situation would require the top plate to be fabricated over a solid layer. As the effect of a different dielectric environment (n>1) would be to modify the evanescent fields and the resonance, a similar effect would require different bar-to-bar distances. In Fig. 4 we present a similar geometry as before but with the bottom bar embedded in SiO$_2$, n=1.5, and when the distance between the bars is D=80nm. It is apparent that the overall physical picture previously discussed remains valid, but there is a red-shift in the location of the resonances due to the higher refractive index of the environment. Naturally, the sweet point also experiments a red-shift, from 1.36μm to 1.43μm.

## 3.  Time-resolved Brillouin scattering enabled by chirality

When considering high-frequency acoustic phonons, that is, with frequencies higher than a few gigahertz, the use of Brillouin scattering [24] and coherent phonon generation [25,26] techniques become the standard experimental characterization tools. [27–29] Let us consider the particular case of an impulsive excitation of coherent acoustic phonons in a metallic bar nanoantenna. [9] In this case, an ultrafast -femtosecond or picosecond- pulsed laser impinges the plasmonic structure. If the incident polarization matches the main axis and the wavelength matches the resonant wavelength of the antenna, then a longitudinal plasmon is excited. This process takes place in the femtosecond timescale. The excited longitudinal plasmon will eventually decay –lose its coherence, thus inducing a rise in temperature with a spatial profile quite similar to that of the plasmon. The decay of the plasmon takes place in the picosecond timescale. The temperature rise has an associated thermal expansion of the antenna that evolves in time as a combination of the time evolutions of the vibrational eigenmodes. Note that the excitation process is intimately related to the initially excited plasmon. If the incident laser pulse has the wrong polarization or wavelength, the plasmon is not excited, and consequently, the coherent phonons are not efficiently generated. [7]

In a bar with a length of approximately 100 nm, the longitudinal acoustic mode frequency will be around 15 GHz, but several other modes can be excited in the system. This frequency range has an associated period of tens of ps. The acoustic phonons induce two effects in the nanoantenna: i) they modulate its shape, and size and ii) they modulate the electronic density and thus the dielectric constant. Both effects result in the modulation of the optical response of the system. By sending a second, delayed in time, ultrafast laser pulse (probe) it is possible to measure the instantaneous optical state of the system. And by tuning the delay between pump and probe pulses and performing multiple experiments, it is possible to reproduce the time evolution of the reflectivity. As with the generation of phonons, the sensitivity of this measurement is also determined by the plasmonic properties of the system. It is a necessary condition that the probe laser couples to a plasmonic mode, and that the spatial profile of this plasmonic mode overlaps with the spatial profile of the considered acoustic phonons. In other words, the probe reflectivity senses only variations of the plasmonic modes to which it is coupled. The only condition imposed on the excitation pulse is that its duration should be shorter than a half period of the studied acoustic modes, to allow an impulsive excitation. Considering fs/ps laser pulses largely complies with this condition. In such a coherent phonon generation-detection experiment, the maximum generation is usually associated with the maximum absorption, while the best condition for detection is at the flank of an optical mode. In trivial optical resonators, such as micropillar microcavities or nanobar antennas, these two conditions -associated with the same optical resonance- cannot be achieved at the same wavelength. In this work, we report on a structure in that absorption and scattering channels can be well separated at the same wavelength by using the appropriate helicity.

As it was explained before, to enhance the signals in this kind of time-resolved Brillouin scattering experiments, two conditions have to be simultaneously fulfilled: i) have an enhanced absorption and minimum scattering for the generation, and ii) present an enhanced scattering cross-section and minimal absorption for the detection of phonons. Two strategies are usually used to reach these conditions, the chromatic separation of pump-probe laser beams with the drawback of using two different ultrafast lasers or to use cross-polarized beams preventing the simultaneous optimization in simple systems.

We propose using the impinging radiation helicity to perform coherent phonon generation experiments in chiral plasmonic structures. We have shown that the structure of the twisted bar structure presents a strong chiral behavior. In the considered system, for a wavelength of 1360 nm, LCP light presents a strong absorption cross-section and attenuated scattering, while for R-polarized light the opposite is true. It results that by using LCP it is possible to have an efficient coherent generation of the acoustic phonons of the structure. In Fig. 5 we show the electromagnetic field intensity ($|E|^2/|E_0|^2$, being $|E_0|^2$ the intensity of the impinging beam) corresponding to LCP incidence at a wavelength of 1360nm. The field is localized in the central part of both bars, and the kinetic energy of the fast electrons will be released uniformly 100 nm around the hot spots. For the sake of simplicity, we can further assume that these phonons will correspond mainly to the breathing modes of the bars, with similar spatial profiles as those previously studied in other works. [7,8]

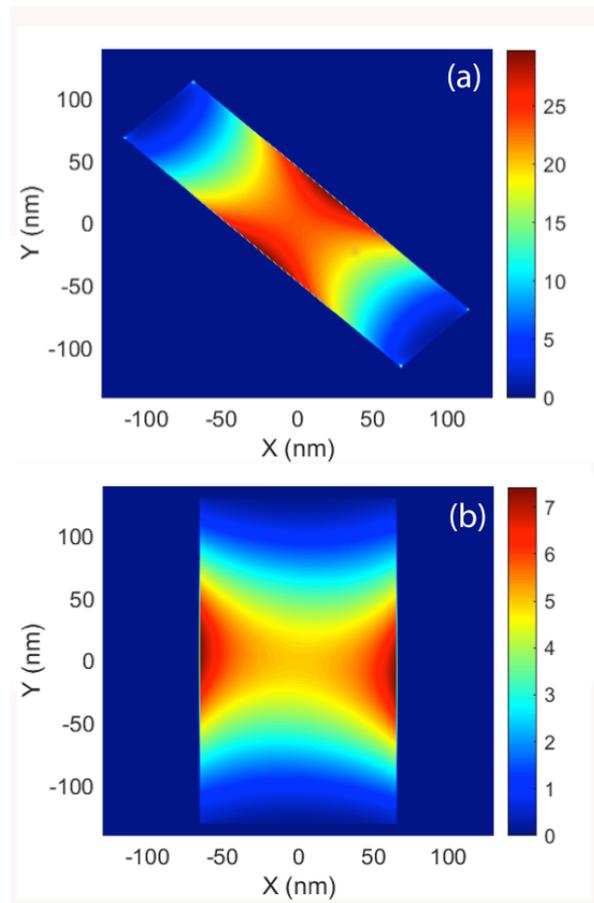

**Fig. 5:** Normalized spatial field intensity at the bottom surface of the top bar (a), and at the top surface of the bottom bar (b). As it can be seen the field intensity closely resembles that of a dipole, but located at the interaction region only.

The fundamental acoustic mode of the bars (See the supplementary information) is a longitudinal mode along the long axis of the bars. In a simplified model, the relevant parameter to characterize the structure sensitivity is the derivative of the scattering cross-section with the size of the bar. In Fig 6 we show the difference of the scattering CS (and absorption, for completeness) for RCP light corresponding to expanded-contracted bars (±2% in the long axis). We can observe that an optimal condition for scattering is achieved at ca.1360 nm. Therefore, at that sweet point, the system presents i) for LCP light there is strong absorption and weak scattering; and ii) for RCP light, the absorption is smaller than for the LCP light, and the scattering cross-section –and its variation under expansion or contraction- bigger than for the LCP light. A simple cross-polarized setup with a synchronous detection system can be easily adapted to work with the helicity, taking full advantage of the chirality of the proposed systems.

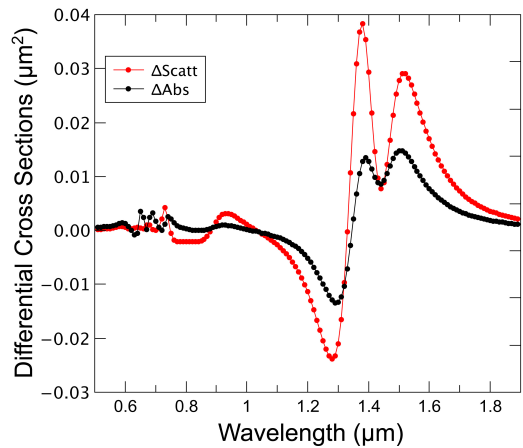

**Fig. 6:** Differential absorption and scattering cross-sections for RCP incidence light assuming an expansion/contraction of 2% over the long axis for both nanoplates nominal geometrical parameters. It can be seen that for scattering the maximum signal appears at the sweet point 1360 nm. At the same point, the differential cross-section for absorption is close to zero.

## 4. Discussion and conclusions

We presented a simple structure, feasible to fabricate, based on crossed elongated bars where light-handedness defines the dominating cross-section absorption or scattering, with a 200% difference from its counterpart (scattering or absorption). The fine-tuning of the interacting distance of the element of the structure allows us to optimize the response of the system to, at a given wavelength, present maximal (minimal) absorption (scattering) cross-section for LCP light, and exactly the opposite for RCP light. The simplicity of the structure leads to ample room for optimization, which can be made via multiparametric variations of the geometry of both interacting nanoplates, the incorporation of an "active substrate" [22,30,31], or modifying the material composition of the nanoplates to include effects such as magneto-optics [32] or hyperbolicity [33].

In nanophononics, [28,34–37] the lack of standard transducers to generate and detect acoustic phonons forced researchers to develop novel experimental schemes based on all-optical techniques. In the most used protocol, [25,26] for a given wavelength, an ideal optoacoustic transducer presents three characteristics: high pump absorption to warrant the impulsive generation of acoustic phonons, and high scattering cross-section of the probe to detect the modification of the optical properties due to the phonons. The use of structures sustaining chiral optical modes appears as a natural, yet unexplored strategy to address these conditions. The proposed chiral system opens the way to enhanced and selective coherent phonon excitation and detection.

Finally, the use of chiral structures like the ones reported in this work could be extended to the domain of quantum applications, such as QD-based single-photon emitters where helicity could become a new design parameter [38,39].

## Acknowledgments

B. C. LdL and A.G.-M. acknowledge support from the Ministerio de Ciencia e Innovación through Grants Nos. TED2021-131417B-I00 (Bigplan-6G), PDC2021-121833-I00 (Raindrops) and PID2019-109905GA-C22 (Symped). B. C. LdL is also grateful to the "Yo Investigo" program of the Comunidad de Madrid. N.D.L.-K. acknowledges funding from the European Research Council Starting Grant No.715939, Nanophennec.

B. Castillo López de Larrinzar[1], C. Xiang[2], E. Cardozo de Oliveira[2],

N. D. Lanzillotti-Kimura[2,*], A. García-Martín[1,+]

# Towards chiral acoustoplasmonics

# Supplementary information

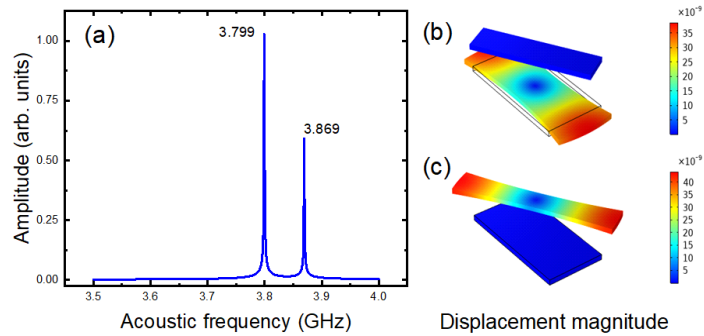

**Figure S1**: (a) Acoustic spectrum of the two twisted nanobars. The acoustic modes of the nanobars are obtained using the finite element method solver COMSOL Multiphysics. A thermal expansion due to a 5 K temperature rise is applied to both bars, which results in a localized strain in the bars. Notice that in this simulation we consider the case of two twisted bars without any solid environment. The frequency domain study calculates the displacements in all directions of the two bars between frequencies of 3.5 to 4 GHz, with a step of 0.001 GHz. Two distinct peaks in the result correspond to the eigenmodes of the two bars. Since the bars are not connected by any solid medium, each one of the modes is fully localized in a single bar. The slight difference in the frequency is due to the lateral confinement.

[1] Instituto de Micro y Nanotecnología IMN-CNM, CSIC, CEI UAM + CSIC, Isaac Newton 8, Tres Cantos, Madrid 28760, Spain
[2] Université Paris-Saclay, CNRS, Centre de Nanosciences et de Nanotechnologies, 10 Boulevard Thomas Gobert, 91120 Palaiseau, France
*email : daniel.kimura@c2n.upsaclay.fr
+email : a.garcia.martin@csic.es

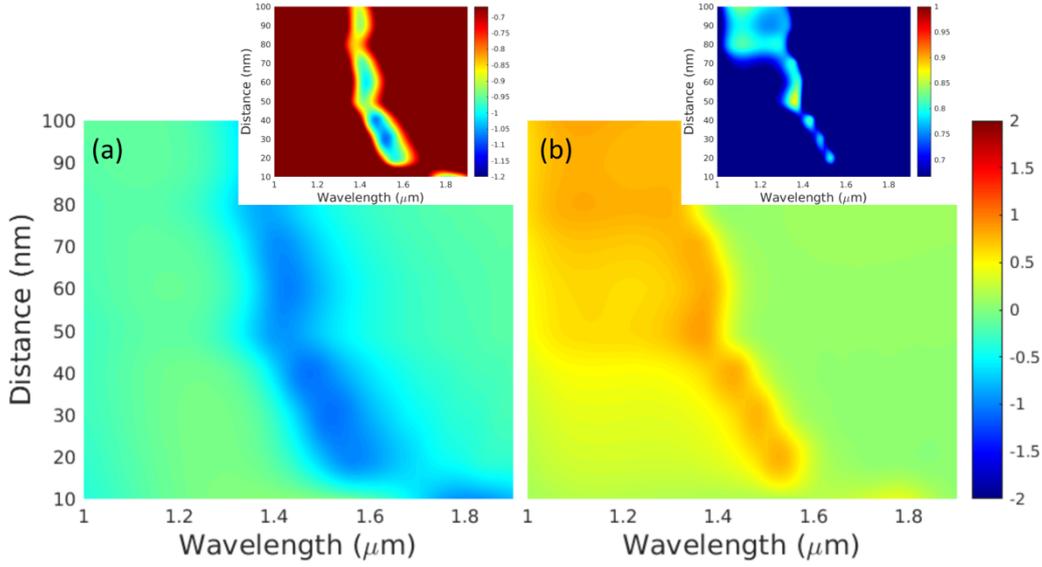

**Figure S2**: Figure of merit of the "asymmetry factor" (g), defined as ($\sigma_{Sc,Ab}^{R,L}$ stands for the Scattering or Absorption cross section for RCP or LCP):

$$g_{Sc,Ab} = 2\frac{\sigma_{Sc,Ab}^{R} - \sigma_{Sc,Ab}^{L}}{\sigma_{Sc,Ab}^{R} + \sigma_{Sc,Ab}^{L}}.$$

The value of g is bound between -2 and 2 for the extreme case where one of them is identically zero. Note that for a situation where one cross-section doubles the other $\sigma^{R} = 2\sigma^{L}$ then g=2/3 and if $2\sigma^{R} = \sigma^{L}$ then g=-2/3. In the spectral region of interest, for absorption (a) the prevalent polarization is LCP, whereas for scattering it is RCP (b), as already grasped from Figs. 2 and 3 in the main text. The larger the separation the shorter the wavelength where the maximum asymmetry occurs. In the insets we have saturated the false color maps to show the regions where the asymmetry is either smaller than -2/3 or larger than 2/3, evidencing ratios between the two polarizations larger than 100%. As depicted the sweet region where both confluence is for distances ca. D=60nm and wavelengths ca. 1360nm.